\documentclass[10pt,conference]{IEEEtran}\IEEEoverridecommandlockouts
\usepackage{epsfig,graphicx,subfigure,psfrag,amsmath,cases,bm}
\usepackage{latexsym,amssymb,amsmath,epsfig,subfigure,algorithm,mathtools}
\usepackage{algorithmic}
\usepackage{color}
\usepackage{url}
\usepackage{scrtime}
\author{\IEEEauthorblockN {Dongfang Xu, Yan Sun, Derrick Wing Kwan Ng, and Robert Schober\vspace*{-13mm}}

\thanks{D. Xu, Y. Sun, and R. Schober are with Friedrich-Alexander-University Erlangen-N\"urnberg, Germany. D. W. K. Ng is with University
of New South Wales, Australia.%
}
}

\title{Robust Resource Allocation for UAV Systems with UAV Jittering and User Location Uncertainty\vspace{-4mm}}

\newtheorem{T-Prob}{Transformed Problem}

\DeclareMathOperator{\mino}{minimize}

\newcommand{\qed}{\hfill \ensuremath{\blacksquare}}

\begin{document}
\maketitle
\begin{abstract}
In this paper, we investigate resource allocation algorithm design for multiuser unmanned aerial vehicle (UAV) communication systems in the presence of UAV jittering and user location uncertainty. In particular, we jointly optimize the two-dimensional position and the downlink beamformer of a fixed-altitude UAV for minimization of the total UAV transmit power. The problem formulation takes into account the quality-of-service requirements of the users, the imperfect knowledge of the antenna array response (AAR) caused by UAV jittering, and the user location uncertainty. Despite the non-convexity of the resulting problem, we solve the problem optimally employing a series of transformations and semidefinite programming relaxation. Our simulation results reveal the dramatic power savings enabled by the proposed robust scheme compared to two baseline schemes. Besides, the robustness of the proposed scheme with respect to imperfect AAR knowledge and user location uncertainty at the UAV is also confirmed.
\end{abstract}
\renewcommand{\baselinestretch}{0.91}
\normalsize
\vspace{-2mm}
\section{Introduction}
\vspace{-1mm}
Recently, wireless communication employing unmanned aerial vehicles (UAVs) has received much attention as a promising approach for offering ubiquitous real-time high data-rate communication services \cite{wong_schober_ng_wang_2017}\nocite{hayat2016survey,7572068,solaruav,7888557}--\cite{secureuav}. Compared to conventional cellular systems, which depend on a fixed terrestrial infrastructure, UAV-assisted communication systems can flexibly deploy UAV-mounted transceivers to a target area to provide on-demand connectivity. For instance, in the case of natural disasters and disease outbreaks, UAVs can be employed as flying base stations to offer temporary communication links in a timely manner. Moreover, benefiting from their high mobility and maneuverability, UAVs are able to adapt their positions and trajectories based on the environment and the terrain which results in extra degrees of freedom for potential system performance improvement \cite{hayat2016survey}. In \cite{7572068}, the authors proposed a new framework for joint power allocation and UAV trajectory optimization to maximize the system throughput of a mobile relaying system. The authors of \cite{solaruav} considered a multicarrier solar-powered UAV communication system and proposed the joint design of power and subcarrier allocation and three-dimensional (3-D) UAV positioning for maximization of the system sum throughput. In \cite{7888557}, the authors studied UAV trajectory design for maximization of the energy-efficiency of a UAV communication system. Besides, secure UAV communications was studied in \cite{secureuav} where the trajectory of a UAV and its transmit power were jointly optimized to maximize the system secrecy rate. However, \cite{7572068}\nocite{solaruav,secureuav,7888557}--\cite{secureuav} assumed that the channel state information (CSI) of the users was perfectly known at the UAV which may not hold in practice. 

In practical UAV communication systems, UAV-mounted transceivers flying in the air may encounter strong wind gusts, which leads to random body jittering with respect to angular movements \cite{choi2015dynamics}. The estimation accuracy of the angle of departure (AoD) between the UAV and the users is impaired by this jittering which results in non-negligible AoD estimation errors \cite{ahmed2010flight}. Moreover, due to the weather conditions and electromagnetic interference, the information about the user location may be imperfect at the UAV \cite{GPSReport}. As a result, additional path loss resulting from user location uncertainty may impair the communication links between the UAV and users. Thus, perfect CSI knowledge of the users cannot be guarantee at the UAV, and the system performance is degraded due to the imperfectness of the CSI \cite{6735741}. On the other hand, multiple antennas performing beamforming can be employed to improve spectral efficiency in future multiuser communication systems. However, the results in \cite{7572068}\nocite{solaruav,secureuav,7888557}--\cite{secureuav} are valid for single-antenna UAVs and are not applicable to multiple-antenna UAVs, since the positioning of the UAV is coupled with the beamformer design. To the best of the authors' knowledge, optimal resource allocation design for multiuser multiple-antenna UAV communication systems in the presence of imperfect CSI has not been investigated in the literature yet.
\par
In this paper, we address the above issues. To this end, the resource allocation algorithm design is formulated as a non-convex optimization problem for minimization of the total transmit power of a downlink (DL) UAV communication system taking into account the quality-of-service (QoS) requirements of the users and imperfect CSI knowledge of the links between the UAV and the users. Thereby, we linearize the antenna array response (AAR) with respect to the AoD estimation errors, since these errors are expected to be small in practice. The formulated non-convex problem is solved optimally by applying transformations and semidefinite programming (SDP) relaxation.
\vspace{-1mm}
\section{System and CSI Models}
\vspace{-1mm}
In this section, we present the system and CSI models for multiuser DL UAV communication. However, first we introduce some notation.
\vspace{-1mm}
\subsection{Notation}
\vspace{-1mm}
In this paper, matrices and vectors are denoted by boldface capital and lower case letters, respectively. $\mathbb{R}^{N\times M}$ and $\mathbb{C}^{N\times M}$ denote the sets of all $N\times M$ real-valued and complex-valued  matrices, respectively. $\mathbb{H}^{N}$ denotes the set of all $N\times N$ Hermitian matrices. $\mathbf{I}_{N}$ denotes the $N-$dimensional identity matrix. $|\cdot|$ and $||\cdot||_2$ represent the absolute value of a complex scalar and the Euclidean norm of a vector, respectively. $\mathbf{x}^T$ and $\mathbf{x}^H$ denote the transpose of vector $\mathbf{x}$ and the conjugate transpose of vector $\mathbf{x}$, respectively. $\mathrm{diag}(a_1, \cdots, a_n)$ denotes a diagonal matrix whose diagonal entries are $a_1, \cdots, a_n$. $\mathrm{Rank}(\mathbf{A})$ and $\mathrm{Tr}(\mathbf{A})$ are the rank and the trace of matrix $\mathbf{A}$, respectively. $\mathbf{A}\succeq\mathbf{0}$ means matrix $\mathbf{A}$ is positive semidefinite. $\mathbf{A}\circ\mathbf{B}$ denotes the Hadamard product of two matrices $\mathbf{A}$ and $\mathbf{B}$ having the same dimensions. $\mathcal{E}\left \{ \cdot \right \}$ denotes statistical expectation. $x\sim \mathcal{CN}(\mu ,\sigma^2)$ indicates that random variable $x$ is circularly symmetric complex Gaussian distributed with mean $\mu$ and variance $\sigma^2$. $\overset{\Delta }{=}$ means ``defined as''. $\nabla_{\mathbf{x}} f(\mathbf{x})$ denotes the gradient vector of function $f(\mathbf{x})$, i.e., its components are the partial derivatives of $f(\mathbf{x})$. $f^{(n)}({a})$ represents the $n$-th order derivative of $f({x})$ at $x=a$.
\vspace{-1mm}
\subsection{Multiuser UAV Communication System}
\vspace{-1mm}
\begin{figure}[t]
\hspace{9mm}\includegraphics[width=2.2in]{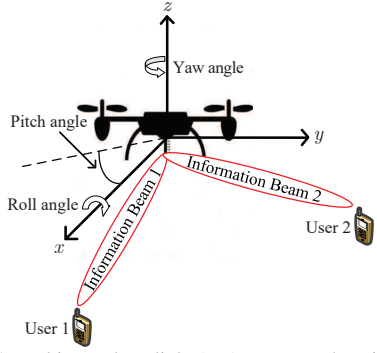} \vspace*{-4mm}
\caption{A multiuser downlink (DL) unmanned aerial vehicle (UAV) communication system with one UAV and $K=2$ users. The three-dimensional Cartesian coordinate system indicates the pitch, yaw, and roll angles of the UAV.
}
\label{fig:UAV-model}\vspace*{-5mm}
\end{figure}\vspace*{-0mm}
\par
The considered multiuser DL UAV communication system model consists of one rotary-wing UAV-mounted transmitter and $K$ users, cf. Figure \ref{fig:UAV-model}. The UAV-mounted transmitter is equipped with ${N_{\mathrm{T}}}$ antenna elements, and the ${N_{\mathrm{T}}}$ antenna elements are equally spaced forming a uniform linear array (ULA). Moreover, the flight height of the UAV is fixed at $z_0$ to avoid obstacles. Besides, we assume that all $K$ users are single-antenna devices. For convenience, we define the set of all users $\mathcal{K}$ as $\mathcal{K}=\left \{ 1,\cdots, K \right \}$.
\par
In each scheduling time slot, the UAV transmits $K$ independent signals simultaneously to the $K$ DL users. Specifically, the transmit signal vector to desired user $k\in\mathcal{K}$ is given by
\vspace{-3mm}
\begin{equation}
\mathbf{x}_k=\mathbf{w}_ks_k,    
\end{equation}
where ${s_k}\in\hspace{-0.5mm} \mathbb{C}$ and $\mathbf{w}_k\in\hspace{-0.5mm} \mathbb{C}^{{\mathit{N}_{\mathrm{T}}}\times 1}$ represent the information symbol for user $k$ and the corresponding beamforming vector, respectively. Without loss of generality, we assume $\mathcal{E}\{\left |s _{k} \right|^2\}=1$, $\forall\mathit{k} \in \mathcal{K}$.
\par
In this paper, we assume that the air-to-ground links between the UAV and the users are line-of-sight (LoS) channels.
In practice, since UAVs fly in the air such that scatterers are encountered with a low probability, the communication links between the UAV and the ground users are typically LoS-dominated \cite{Lin2018TheSI}. In particular, the channel vector between the UAV and user $k$ is modelled as \cite{sun2017air}
\vspace{-2mm}
\begin{equation}
\mathbf{h}_k=\sqrt{\varrho}\left\|\mathbf{r}_0- \mathbf{r}_k \right \|_2^{-1}\mathbf{a}(\theta_k),\label{channelvec}
\end{equation}
where $\varrho=(\frac{\lambda_c}{4\pi})^2$ is a constant with $\lambda_c$ being the wavelength of the center frequency of the carrier. $\mathbf{r}_{0}=(x_{0},~y_{0},~z_{0})$ and $\mathbf{r}_{k}=(x_{k},~y_{k},~0)$ denote the 3-D Cartesian coordinates of the UAV and user $k$, respectively. Moreover, $\sqrt{\varrho} \left\|\mathbf{r}_0- \mathbf{r}_k \right \|^{-1}_2$ is the average channel power gain between the UAV and user $k$. Besides, $\mathbf{a}(\theta_k)$ represents the AAR between the UAV and user $k$ and is given by \cite{Tse:2005:FWC:1111206}
\vspace{-3mm}
\begin{equation}
\mathbf{a}(\theta_k)=\begin{pmatrix}
1,e^{-j2\pi\frac{b}{\lambda _c}\mathrm{cos}\theta_k}, \hdots,e^{-j2\pi\frac{b}{\lambda _c}(N_\mathrm{T}-1)\mathrm{cos}\theta_k}
\end{pmatrix}^T,\label{steeringvec}
\end{equation}
where $\theta_k$ is the AoD of the path between the ULA and DL user $k$, and $b$ is the separation distance between the antennas equipped at the ULA.
\par
\begin{figure}[t]
\hspace{4mm}\includegraphics[width=2.8in]{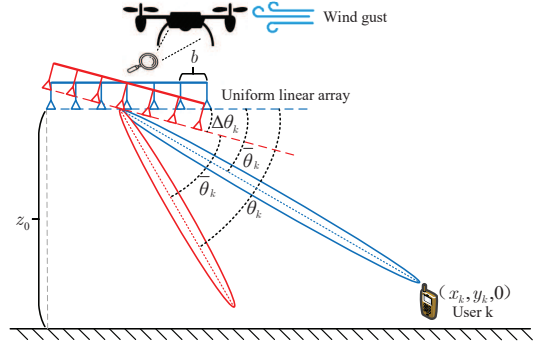} \vspace*{-3mm}
\caption{DL channel model assuming line-of-sight between each antenna element and DL user $k$. The blue beam points to the desired user $k$, and the red beam shows the actual beam direction impaired by a wind gust. $\overline{\theta}_k$, $\theta_k$, and $\Delta\theta_k$ denote the estimated angle of departure (AoD), the actual AoD, and varying pitch angle, respectively.
}
\label{fig: channel model}\vspace{-4mm}
\end{figure}
\par
Therefore, the received signal at user $\mathit{k} \in \mathcal{K}$ is given by
\vspace{-2.5mm}
\begin{equation}
{\mathit{d_k}} = \underset{\mathrm{desired~signal}}{\underbrace{{\mathbf{h_\mathit{k}^\mathit{H}}}\mathbf{w_\mathit{k}}s_k}} +  \underset{\mathrm{multiuser~interference}}{\underbrace{\underset{ r\in\mathcal{K}\setminus \left \{ k \right \}}{\sum}\mathbf{h_\mathit{k}^\mathit{H}}\mathbf{w_\mathit{r}}s_r}} + n_k,
\end{equation}
where $n_k$ captures the joint effect of the background noise and the thermal noise at the receive antenna of user $k$. We model $n_k$ as additive complex Gaussian noise with variance $\sigma ^2_{n_k}$, i.e., $n_k\sim \mathcal{CN}(0,\sigma ^2_{n_k})$.

\vspace{-1mm}
\subsection{Channel State Information Uncertainty}
\vspace{-1mm}
In practice, the stability of the UAV in the air is affected by the random nature of wind gusts \cite{6350400}. In particular, UAVs suffer from body jittering in the presence of strong wind, and the flight behaviour of the UAV changes with respect to the pitch, yaw, and roll angles \cite{da2017advanced}, cf. Figure \ref{fig:UAV-model}. Moreover, varying pitch angles capture the main impact of UAV jittering, since horizontal wind gusts in the lower troposphere are the dominant cause of UAV jittering \cite{martin2011meteorological}. As a result, the estimation of the AoD $\theta_k$ is influenced by the varying pitch angle. In fact, due to the randomness of wind gusts, the onboard sensors of the UAV may not be able to measure the exact pitch angle. Hence, AoD estimation errors occur which leads to imperfect AoD knowledge at the UAV. To capture this effect, we adopt a deterministic model for the resulting AoD uncertainty \cite{7111366}. Specifically, the AoD between the ULA and DL user $k$, i.e., $\theta_k$, is modelled as
\vspace{-1mm}
\begin{equation}
\theta_k= \overline{\theta}_k+\Delta \theta_k,~\Omega _k=\left \{ \Delta \theta_k\in \mathbb{R}| \left | \Delta \theta_k\right |\leq \alpha\right \}, ~\forall k\in\mathcal{K},\label{uncertaintyset}
\end{equation}
where $\overline{\theta}_k$ and $\Delta \theta_k$ represent the estimated AoD between the ULA and user $k$ and the unknown AoD uncertainty, respectively, cf. Figure \ref{fig: channel model}. Besides, the continuous set $\Omega _k$ contains all possible AoD uncertainties with bounded maximum pitch variation $\alpha$\footnote{In practice, the pitch angle varies between $10^{-1}$ rad to $10^{-3}$ rad \cite{ahmed2010flight}.}. In practice, the value of $\alpha$ depends on the climatic conditions and the UAV model \cite{yeo2016onboard}.
Then, the imperfect AAR is given by
\vspace{-1mm}
\begin{equation}
\hspace{-1mm}\mathbf{a}(\theta_k)\hspace{-1.2mm}=\hspace{-1.2mm}\left ( 
\hspace{-0.5mm}1,\hspace{-0.5mm}e^{-j2\pi\frac{b}{\lambda _c}\mathrm{cos}(\overline{\theta}_k\hspace{-0.5mm}+\Delta \theta_k)}\hspace{-0.5mm},\hspace{-0.5mm}\hdots,\hspace{-0.5mm}e^{-j2\pi\frac{b}{\lambda _c}(\hspace{-0.5mm}N_\mathrm{T}-\hspace{-0.5mm}1\hspace{-0.5mm})\mathrm{cos}(\overline{\theta}_k\hspace{-0.5mm}+\Delta \theta_k)\hspace{-1mm}}\right )^{\hspace{-0.5mm}T}\hspace{-2mm}.\label{steeringvec2}
\end{equation}
We note that $\mathbf{a}(\theta_k)$ is a nonlinear function with respect to $\Delta \theta_k$, which complicates robust resource allocation algorithm design. To tackle this problem, and since the $\Delta \theta_k$ are generally small, for a given $\overline{\theta} _k$, we approximate $\mathbf{a}(\theta _k)$ by applying a first order Taylor series expansion:
\begin{equation}
\mathbf{a}(\theta _k)\approx \mathbf{a}^{\hspace{-0.5mm}(\hspace{-0.2mm}0\hspace{-0.2mm})}(\overline{\theta} _k)+\mathbf{a}^{\hspace{-0.5mm}(\hspace{-0.2mm}1\hspace{-0.2mm})}(\overline{\theta} _k)(\theta _k-\overline{\theta} _k),\label{smeq}
\end{equation}
\vspace{-3mm}where 
\begin{eqnarray}
&&\hspace{-12mm}\mathbf{a}^{\hspace{-0.5mm}(\hspace{-0.2mm}0\hspace{-0.2mm})}(\overline{\theta} _k)\hspace{-1mm}=\hspace{-1mm}\begin{pmatrix}1,e^{-j2\pi\frac{b}{\lambda _c}\mathrm{cos}\overline{\theta}_k}, \hdots,e^{-j2\pi\frac{b}{\lambda _c}(N_\mathrm{T}-1)\mathrm{cos}\overline{\theta}_k}\end{pmatrix}^{\hspace{-0.5mm}T}\hspace{-0.5mm},\\
&&\hspace{-12mm}\mathbf{a}^{\hspace{-0.5mm}(\hspace{-0.2mm}1\hspace{-0.2mm})}\hspace{-0.5mm}(\overline{\theta} _k)\hspace{-1mm}=\hspace{-1mm}\small\begin{pmatrix}
\hspace{-0.2mm}0, 
\hspace{-0.5mm}j2\pi\frac{b}{\lambda_c}\mathrm{sin}\overline{\theta}_k, 
\hspace{-0.5mm}\hdots,\hspace{-0.5mm}j2\pi\left (\hspace{-0.5mm} N_\mathrm{T}\hspace{-0.5mm}-\hspace{-0.5mm}1 \hspace{-0.5mm}\right )\frac{b}{\lambda_c}\mathrm{sin}\overline{\theta}_k
\end{pmatrix}^{\hspace{-0.5mm}T}\hspace{-2.2mm}\circ\hspace{-0.5mm}\mathbf{a}^{\hspace{-0.3mm}(\hspace{-0.2mm}0\hspace{-0.2mm})}\hspace{-0.5mm}(\overline{\theta} _k).
\end{eqnarray}
Then, the AAR between the UAV and user $k$ is modeled as
\vspace{-3mm}
\begin{equation}
\mathbf{a}_k=\mathbf{\overline{a}}_k+\Delta \mathbf{a}_k,\label{newsteeringvec}    
\end{equation}
where $\mathbf{\overline{a}}_k$ and $\Delta \mathbf{a}_k\in \mathbb{C}^{N_\mathrm{T}\times 1}$ are defined as
\vspace{-2.5mm}
\begin{equation}
\mathbf{\overline{a}}_k\overset{\Delta }{=}\mathbf{a}^{(0)}(\overline{\theta} _k)\textrm{~and~} \Delta \mathbf{a}_k\overset{\Delta }{=}\mathbf{a}^{(1)}(\overline{\theta} _k)\Delta \theta_k,\label{steeringvec3}
\end{equation}
respectively. We note that $\mathbf{\overline{a}}_k$ and $\Delta \mathbf{a}_k$ are the AAR estimate of user $k$ and the corresponding linearized AAR uncertainty, respectively.
\par
\textit{Remark 1:~}We note that the linearized AAR model in (\ref{smeq}) is employed since $\Delta \theta_k$ is small in practice and to make resource allocation design tractable. In our simulations, we adopt the nonlinear AAR model in (\ref{steeringvec2}) to evaluate the proposed resource allocation algorithm. 
\par
On the other hand, the user location information at the UAV, provided e.g. by GPS \cite{whitepaper}, may be also imperfect due to radio signal interference, satellite shadowing, and atmospheric impairments\footnote{In practice, positioning errors in forth-generation long-term evolution (4G LTE) networks are typically in the range from 10 meters to 200 meters, depending on the adopted positioning protocol \cite{whitepaper}.}. Thus, in this paper, we also take into account the user location uncertainty for robust resource allocation algorithm design. Specifically, since we assume that all users are on the ground, their $z$ coordinates are all set to 0. Then, the $x-y$ coordinates of user $k$ are modelled as
\vspace{-1mm}
\begin{equation}
x_k=\overline{x}_k+\Delta x_k,~~y_k=\overline{y}_k+\Delta y_k,
\end{equation}
respectively, where $\overline{x}_k$ and $\overline{y}_k$ are the user location estimates available at the UAV, and $\Delta x_k$ and $\Delta y_k$ denote the respective location uncertainties. Furthermore, we assume that the UAV knows its own location perfectly. In fact, thanks to onboard multi-sensor systems and advanced positioning strategies for UAVs, the positioning accuracy of UAVs can be improved to centimeter level \cite{zimmermann2017precise}. To simplify notation, we define
\begin{eqnarray}
\label{comp}
&&{\mathbf{r}}'_0=(x_0,~y_0)^T,~{\mathbf{r}}'_k=(x_k,~y_k)^T,\\
&&{\overline{\mathbf{r}}}'_k= (\overline{x}_k,~\overline{y}_k)^T,~{\Delta{\mathbf{r}}}'_k=(\Delta x_k,~\Delta y_k)^T,
\end{eqnarray}
where vectors ${\mathbf{r}}'_0$, ${\mathbf{r}}'_k$, ${\overline{\mathbf{r}}}'_k$, and ${\Delta{\mathbf{r}}}'_k$ include the $x-y$ coordinates of the UAV, the actual $x-y$ coordinates of user $k$, the estimated $x-y$ coordinates of user $k$, and the $x-y$ uncertainties of user $k$, respectively. Then, the 3-D Cartesian coordinates of the UAV and user $k$ can be expressed equivalently as 
\begin{eqnarray}
\hspace{4mm}\mathbf{r}_0=
(({{\mathbf{r}}'_0})^T,~z_0)^T
~\mbox{and}~
\mathbf{r}_k=
(({{\overline{\mathbf{r}}}_k'})^T+({{\Delta{\mathbf{r}}}_k'})^T,~ 0)^T,
\end{eqnarray}
and the distance between the UAV and user $k$ can be rewritten as
\begin{equation}
\left \| \mathbf{r}_0- \mathbf{r}_k\right \|_2=\sqrt{\left \| {\mathbf{r}}'_0- ({\overline{\mathbf{r}}}'_k+{\Delta{\mathbf{r}}}'_k)\right \|_2^2+z_0^2}. \label{locationun1}   
\end{equation}
Besides, we define set $\Psi _k$ to collect all possible location uncertainties of user $k$ as follows
\begin{equation}
   \Psi _k\overset{\Delta }{=}\left \{ {\Delta{\mathbf{r}}}'_k\in \mathbb{R}^2|({\Delta{\mathbf{r}}}'_k)^T{\Delta{\mathbf{r}}}'_k\leq D_k^2\right \},~\forall k\in\mathcal{K},\label{locationun2}
\end{equation}
where $D_k$ is the radius of the circular uncertainty region, whose value depends on the positioning accuracy. 

\vspace{-1mm}
\section{Problem Formulation and Solution}
In this section, we formulate the joint power and two-dimensional (2-D) positioning optimization problem for the considered UAV communication system after defining the adopted system performance metric. Then, we solve the resulting problem optimally via SDP relaxation.
\vspace{-1mm}
\subsection{Problem Formulation}
\vspace{-1mm}
The received signal-to-interference-plus-noise ratio (SINR) of user $k$ is given by
\vspace{-2mm}
\begin{equation}
\Gamma _k=\frac{\frac{\varrho}{\left\|\mathbf{r}_0- \mathbf{r}_k \right \|_2^2}\left |{\mathbf{a^\mathit{H}_\mathit{k}}}\mathbf{w}_k\right |^2}{\frac{\varrho}{\left\|\mathbf{r}_0- \mathbf{r}_k \right \|_2^2}\underset{ r\in\mathcal{K}\setminus \left \{ k \right \}}{\sum}{\left |{\mathbf{a^\mathit{H}_\mathit{k}}}\mathbf{w}_r\right |^2}+  \sigma^2_{n_k}}.
\end{equation}
In practice, the endurance of the UAVs is restricted by the limited onboard battery capacity \cite{Gupta2016SurveyOI}. Thus, power-efficient communication for minimization of the UAV transmit power is of utmost importance for UAV-assisted communication systems. Hence, in this paper, we aim to minimize the total UAV transmit power while meeting the QoS requirements of all DL users. The optimal power allocation and 2-D positioning policy for the UAV can be obtained by solving the following optimization problem:
\vspace{-1mm}
\begin{eqnarray}
\label{prob1}
&&\hspace*{1mm}\underset{\mathbf{w}_{\mathit{k}},{\mathbf{r}}'_0}{\mino} \,\, \,\, \underset{ k\in\mathcal{K}}{\sum} \mathbf{w}_{\mathit{k}}^H\mathbf{w}_{\mathit{k}} \\
\notag\mbox{s.t.}\hspace*{1mm}
&&\hspace{-7mm}\mbox{C1:}\left [  \underset{k\in\mathcal{K}}{\sum} \mathbf{w}_k\mathbf{w}_k^H\right ]_{i,i} \leq \mathit{P}_i,~\forall i,\notag\\
&&\hspace{-7mm}\mbox{C2: }
\hspace{-2mm}\underset{{\scriptsize\begin{matrix}\Delta \mathrm{\theta}_k\in\Omega_k,\\ {\Delta{\mathbf{r}}}'_k\in\Psi _k,\\ k\in\mathcal{K}\end{matrix}} }{\mathrm{min}}{\frac{\frac{\varrho}{\left\|\mathbf{r}_0- \mathbf{r}_k \right \|_2^2}\left |{\mathbf{a_\mathit{k}}^H}\mathbf{w}_k\right |^2}{\frac{\varrho}{\left\|\mathbf{r}_0- \mathbf{r}_k \right \|_2^2}\underset{ r\in\mathcal{K}\setminus \left \{ k \right \}}{\sum}{\left |{\mathbf{a_\mathit{k}}^H}\mathbf{w}_r\right |^2}+  \sigma^2_{n_k}}}\geq \Gamma_{\mathrm{req}_k},\notag
\end{eqnarray}
where $\left [ \cdot \right ]_{i,i}$ denotes the $(i,i)$-entry of a matrix. Constraint C1 constrains the transmit power of the $i$-th antenna element of the UAV to not exceed the maximum power allowance $P_i$. In practice, the transmit power of each antenna element is limited individually by the corresponding power
amplifier in the analog front-end. Constraint C2 ensures that the QoS requirements of all users are satisfied, and $\Gamma_{\mathrm{req}_k}$ is the minimum SINR required by user $k$ for reliable information decoding. 
\par
We note that the optimization problem in \eqref{prob1} is non-convex because of the non-convexity of constraint C2. General systematic methods for solving non-convex optimization
problems are not known. In addition, constraint C2 involves infinitely many inequality constraints which makes robust resource allocation algorithm design intractable. However, in the next subsection, we will show that the resulting problem can be solved optimally via SDP relaxation.

\vspace{-1mm}
\subsection{Solution of the Optimization Problem}
\vspace{-1mm}
In this subsection, the problem in \eqref{prob1} is reformulated into an  equivalent form, and then the semi-infinite constraint C2 is transformed into linear matrix inequality (LMI) constraints. Finally, we employ SDP relaxation to recast the considered problem as a convex optimization problem, which allows us to solve it optimally in an efficient manner.
\par
To facilitate SDP relaxation, we define $\mathbf{W_\mathit{k}}=\mathbf{w_\mathit{k}}\mathbf{w_\mathit{k}^\mathit{H}}$, $\mathbf{A_\mathit{k}}= \mathbf{a_\mathit{k}}\mathbf{a_\mathit{k}^\mathit{H}}$, $\forall k\in\mathcal{K}$, and rewrite the problem in (\ref{prob1}) as
\vspace{-1mm}
\begin{eqnarray}
\label{prob2}
&&\hspace{4mm}\underset{\mathbf{w}_{\mathit{k}},{\mathbf{r}}'_0}{\mino} \,\, \,\, \underset{ k\in\mathcal{K}}{\sum} \mathrm{Tr}({\mathbf{W_\mathit{k}}})\\
\notag\mbox{s.t.}\hspace*{1mm}\vspace*{-3mm}
&&\hspace*{-7mm}\mbox{C1:} \left [ \underset{ k\in\mathcal{K}}{\sum}\mathbf{W}_k\right ]_{i,i} \leq \mathit{P}_i,~\forall i,\notag\\
&&\hspace*{-7mm}\mbox{C2: }
\hspace{-1mm}\underset{{\scriptsize\begin{matrix}\Delta \mathrm{\theta}_k\hspace{-0.5mm}\in\hspace{-0.5mm}\Omega_k,\\ {\Delta{\mathbf{r}}}'_k\hspace{-0.5mm}\in\hspace{-0.5mm}\Psi _k,\\ k\in\mathcal{K}\end{matrix}} }{\mathrm{min}}{\hspace{-0.5mm}\frac{\frac{\varrho}{\left\|\mathbf{r}_0- \mathbf{r}_k \right \|_2^2}\mathrm{Tr}({\mathbf{W_\mathit{k}}}\mathbf{A_\mathit{k}})}{\frac{\varrho}{\left\|\mathbf{r}_0- \mathbf{r}_k \right \|_2^2}\underset{ r\in\mathcal{K}\setminus \left \{ k \right \}}{\sum}\hspace{-1mm}\mathrm{Tr}({\mathbf{W_\mathit{r}}}\mathbf{A_\mathit{k}})+  \sigma^2_{n_k}}}\hspace{-1mm}\geq\hspace{-1mm} \Gamma_{\mathrm{req}_k},\notag\\
&&\hspace*{-7mm}\mbox{C3:~}{\mathbf{W_\mathit{k}}}\succeq \mathbf{0},~\forall k,~~~~\mbox{C4:~}\mathrm{Rank}({\mathbf{W_\mathit{k}}})\leq 1,~\forall k,\notag
\end{eqnarray}
where ${\mathbf{W_\mathit{k}}}\succeq \mathbf{0}$, $\mathbf{W_\mathit{k}}\in\mathbb{H}^{N_{\mathrm{T}}}$, and $\mathrm{Rank}({\mathbf{W_\mathit{k}}})\leq 1$ in constraints C3 and C4 are imposed to ensure that $\mathbf{W_\mathit{k}} = \mathbf{w_\mathit{k}}\mathbf{w_\mathit{k}^\mathit{H}}$ holds after optimization. Moreover, we note that constraint C2 is a semi-infinite constraint, as the coupled uncertainty variables $\Delta \mathrm{\theta}_k$ and ${\Delta{\mathbf{r}}}'_k$ are continuous in the sets $\Omega_k$ and $\Psi _k$, respectively. In order to transform constraint C2 into a tractable form, we first decouple the uncertainty variables by multiplying simultaneously the numerator and the denominator of the fractional term with $\varrho^{-1}\left\|\mathbf{r}_0- \mathbf{r}_k \right \|_2^2$. Then, we introduce a scalar slack variable $\eta_k$, and rewrite constraint C2 equivalently as
\vspace{-3mm}
\begin{eqnarray}
&&\hspace{-13mm}\mbox{C2a:\hspace{1mm}}\mathrm{Tr}({\mathbf{W_\mathit{k}}}\mathbf{A_\mathit{k}})\hspace{-1mm}-\Gamma _{\mathrm{req}_k}\hspace{-3mm}\underset{ r\in\mathcal{K}\setminus \left \{ k \right \}}{\sum}\hspace{-3mm}\mathrm{Tr}({\mathbf{W_\mathit{r}}}\mathbf{A_\mathit{k}})\hspace{-1mm}\geq\hspace{-1mm} \eta _k, \forall \Delta \mathrm{\theta}_k\hspace{-1mm}\in\hspace{-1mm}\Omega_k,\label{C2a}\\
&&\hspace{-13mm}\mbox{C2b:\hspace{1mm}}\eta _k\geq \Gamma _{\mathrm{req}_k}\frac{\sigma _{n_k}^2\left\|\mathbf{r}_0- \mathbf{r}_k \right \|_2^2}{\varrho}, ~\forall{\Delta{\mathbf{r}}}'_k\in\Psi _k,~\forall k\label{C2b}.
\end{eqnarray}
Next, we introduce a lemma which can be used to transform constraints C2a and C2b into LMIs with a finite number of constraints.
\\
\textit{Lemma~1~(S-Procedure \cite{boyd2004convex}):} Let a function $f_m(\mathbf{x})$, $m\in \left \{ 1,2 \right \}$, $\mathbf{x}\in \mathbb{C}^{N\times 1}$, be defined as
\vspace{-2mm}
\begin{equation}
f_m(\mathbf{x})= \mathbf{x}^H\mathbf{B}_m\mathbf{x}+2\mathrm{Re}\left \{\mathbf{b}^H_m\mathbf{x}  \right \}+c_m,
\end{equation}
where $\mathbf{B}_m\in \mathbb{H}^N$, $\mathbf{b}_m\in \mathbb{C}^{N\times 1}$, and $\mathrm{c}_m\in \mathbb{R}^{1\times 1}$. Then, the implication $f_1(\mathbf{x})\leq0 \Rightarrow f_2(\mathbf{x})\leq0$ holds if and only if there exists a $\delta \geq 0$ such that
\vspace{-2mm}
\begin{equation}
\delta \begin{bmatrix}
\mathbf{B}_1 &  \mathbf{b}_1\\
\mathbf{b}_1^H &  \mathit{c}_1
\end{bmatrix}-\begin{bmatrix}
\mathbf{B}_2 &  \mathbf{b}_2\\
\mathbf{b}_2^H &  \mathit{c}_2
\end{bmatrix}\succeq \mathbf{0},
\end{equation}
provided that there exists a point $\widehat{\mathbf{x}}$ such that $f_m(\widehat{\mathbf{x}})<0$.
\par
By applying (\hspace{-0.3mm}\ref{newsteeringvec}\hspace{-0.1mm}) and (\hspace{-0.3mm}\ref{steeringvec3}\hspace{-0.1mm}), we can rewrite constraint C2a as
\vspace{-5.5mm}
\begin{eqnarray}
\mbox{C2a:~}0\geq&&\hspace{-7mm}\left(\Delta\theta_k\right )^2 \left [\mathbf{a}^{(1)}(\overline{\theta} _k)  \right ]^{\hspace{-0.5mm}\mathit{H}}\hspace{-1mm}(\Gamma _{\mathrm{req}_k}\hspace{-3mm}\underset{ r\in\mathcal{K}\setminus \left \{ k \right \}}{\sum}\hspace{-3mm}\mathbf{W_\mathit{r}}-\mathbf{W_\mathit{k}})\mathbf{a}^{(1)}(\overline{\theta} _k)\notag\\+&&\hspace{-7mm}2 (\Delta\theta_k)\mathrm{Re}\left \{\mathbf{\overline{a}_\mathit{k}^\mathit{H}}( \Gamma _{\mathrm{req}_k}\hspace{-3mm}\underset{ r\in\mathcal{K}\setminus \left \{ k \right \}}{\sum}\hspace{-3mm}\mathbf{W_\mathit{r}}-\mathbf{W_\mathit{k}})\mathbf{a}^{(1)}(\overline{\theta} _k)\right \}\notag\\+&&\hspace{-7mm}\mathbf{\overline{a}_\mathit{k}^\mathit{H}}\left ( \Gamma _{\mathrm{req}_k}\hspace{-3mm}\underset{ r\in\mathcal{K}\setminus \left \{ k \right \}}{\sum}\hspace{-3mm}\mathbf{W_\mathit{r}}-\mathbf{W_\mathit{k}}\right )\mathbf{\overline{a}_\mathit{k}}+\eta_k.\label{C2a-}
\end{eqnarray}
Using Lemma 1, the following implication can be obtained: $\left (  \Delta\theta_k\right )^2-\alpha ^2 \leq 0 \Rightarrow$ C2a holds if and only if there exist $\delta _{k}\geq 0$ such that,
\vspace{-1.5mm}
\begin{eqnarray}
&&\hspace{-6mm}\overline{\mbox{C2}}\mbox{a}\hspace{-0.5mm}:\hspace{-0.5mm}\mathbf{S}_{\overline{\mathrm{C}2}\mathrm{a}_{k}}(\mathbf{W_\mathit{k}},\mathit{\eta_\mathit{k}},\delta_{k})\\
&&\hspace{-6mm}=\hspace{-1mm}\begin{bmatrix}
\delta_{k} \hspace{-2mm}& 0\notag\\
0 \hspace{-2mm}& -\delta_{k}\alpha^2\hspace{-0.5mm}-\hspace{-0.5mm}\eta_k\end{bmatrix}\hspace{-1.5mm}+\hspace{-1mm}\mathbf{U^{\mathit{H}}_{a_{\mathit{k}}}}[\mathbf{W_\mathit{k}}\hspace{-1mm}-\Gamma _{\mathrm{req}_k}\hspace{-3mm}\underset{ r\in\mathcal{K}\setminus \left \{ k \right \}}{\sum}\hspace{-2.5mm}\mathbf{W_\mathit{r}}]\mathbf{U_{a_{\mathit{k}}}}\hspace{-0.5mm}\succeq\hspace{-0.5mm} \mathbf{0},\forall k,\label{SC2a}
\end{eqnarray}
holds, where $\mathbf{U_{a_{\mathit{k}}}}=\left [ \mathbf{a}^{(1)}(\overline{\theta} _k)~\mathbf{\overline{a}_\mathit{k}} \right ]$.
Similarly, based on (\ref{locationun1}) and (\ref{locationun2}), constraint C2b can be rewritten as
\vspace{-2mm}
\begin{eqnarray}
\hspace{-12mm}\mbox{C2b:~}0&&\hspace{-6mm}\geq ({\Delta{\mathbf{r}}}'_k)^\mathit{T}{\Delta{\mathbf{r}}}'_k+2\mathrm{Re}\left \{({\overline{\mathbf{r}}}'_k-{\mathbf{r}}'_0)^\mathit{T}{\Delta{\mathbf{r}}}'_k\right \}\notag\\&&\hspace{-6mm}+~({\overline{\mathbf{r}}}'_k-{\mathbf{r}}'_0)^\mathit{T}({\overline{\mathbf{r}}}'_k-{\mathbf{r}}'_0)+z_0^2-\frac{\varrho}{\Gamma _{\mathrm{req}_k}\sigma _{n_k}^2}\eta_k.\label{C2b-}
\end{eqnarray}
We apply Lemma 1 to C2b and obtain an equivalent LMI constraint:
\vspace{-2mm}
\begin{eqnarray}
&&\hspace{-6mm}\overline{\mbox{C2}}\mbox{b}:\mathbf{S}_{\overline{\mathrm{C}2}\mathrm{b}_{k}}({\mathbf{r}}'_0,\mathit{\eta_\mathit{k}},\mu_{k})\\
&&\hspace{-6mm}=\hspace{-1.5mm}\begin{bmatrix}
\hspace{-0.5mm}(\mu _k-1)\mathbf{I}_2 \hspace{-2mm} & {\mathbf{r}}'_0-{\overline{\mathbf{r}}}'_k\notag\\ 
\hspace{-0.5mm}({\mathbf{r}}'_0-{\overline{\mathbf{r}}}'_k)^T \hspace{-2mm}& -\mu _kD_k^2\hspace{-1mm}-\hspace{-1mm}\left \|  {\overline{\mathbf{r}}}'_k-{\mathbf{r}}'_0\right \|_2^2\hspace{-1mm}-z_0^2+\hspace{-1mm}\frac{\varrho \eta_k}{\Gamma _{\mathrm{req}_k}\sigma _{n_k}^2 }
\end{bmatrix}\hspace{-1.5mm}\succeq\mathbf{0},\forall k,\label{SC2b}
\end{eqnarray}
where $\mu_{k}\geq0$.
\par
Now, the only obstacle to solving problem \eqref{prob2} efficiently is the rank-one constraint $\mbox{C4}$. To handle this problem, we employ SDP relaxation by removing constraint $\mbox{C4}$, and the considered problem becomes an SDP which is given by 
\vspace{-2mm}
\begin{equation}
\underset{\mathbf{W}_{\mathit{k}}\in\mathbb{H}^{N_\mathit{T}},{\mathbf{r}}'_0,\mathit{\eta_\mathit{k}},\delta_{k},\mu_{k}}{\mathrm{minimize}}~\underset{ k\in\mathcal{K}}{\sum} \mathrm{Tr}(\mathbf{W}_{\mathit{k}})\label{OP3}
\end{equation}
\vspace{-4mm}
\begin{align*}
\mathrm{s.t.}~~\mbox{C1,~}{\overline{\mbox{C2}}\mbox{a},~}{\overline{\mbox{C2}}\mbox{b},~}\mbox{C3.}
\end{align*}
The convex problem in \eqref{OP3} can be efficiently solved by standard convex solvers such as CVX \cite{grant2008cvx}. Besides, the tightness of the SDP relaxation is revealed in the following theorem.
\par
\textit{Theorem 1:~}If $\Gamma _{\mathrm{req}_k}>0$, an optimal rank-one beamforming matrix $\mathbf{W}_k$ in \eqref{OP3} can always be obtained.
\par
\textit{Proof:~}Please refer to the Appendix. \qed
\par
Theorem 1 unveils that the optimal beamforming matrix, $\mathbf{W}_k$, is rank-one, and hence, allows the extraction of the optimal beamforming vector $\mathbf{w}_k$, despite the imperfect AoD knowledge and the user location uncertainty at the UAV. 
\par
\textit{Remark 2:~}In this paper, to make the resource allocation design tractable, we design the beamforming vectors for the linearized AAR model in (\ref{newsteeringvec}). However, this approximation may lead to a violation of the QoS constraint $\mbox{C2}$ for the actual nonlinear AAR model in (\ref{steeringvec2}). To circumvent this problem, we solve (\ref{OP3}) for slightly higher minimum required SINRs $\Gamma _{\mathrm{req}_k}+\gamma$, where $\gamma$ is a small positive number, which is chosen sufficiently large to ensure that $\mbox{C2}$ is also met for the nonlinear AAR model.
\begin{table}[t]\caption{System Parameters}\vspace{-3mm}\label{tab:para} \centering
\begin{tabular}{|l|c|}\hline
\hspace*{-1mm}Carrier center frequency and bandwidth & $2.4$ GHz and $200$ kHz\\
\hline
\hspace*{-1mm}ULA antenna element separation, $b$ &  \mbox{$6.25\times10^{-2}$ meter}\\
\hline
\hspace*{-1mm}UAV fixed flight altitude, $z_0$ & $100$ meters \\
\hline
\hspace*{-1mm}DL user noise power, $\sigma^2_{n_k}$&   \mbox{$-110$ dBm} \\
\hline
\hspace*{-1mm}UAV maximum per-antenna transmit power, $P_i$ &  \mbox{$20$ dBm} \\
\hline
\hspace*{-1mm}Minimum required SINR at user $k$, $\Gamma_{\mathrm{req}_k}$ &\mbox{$10$ dB} \\
\hline
\end{tabular}\vspace{-6mm}
\end{table}
\section{Simulation Results}
\vspace{-1.5mm}
In this section, the performance of the proposed resource allocation scheme is investigated via simulations. The simulation parameters are listed in Table \ref{tab:para}. Specifically, there are $K$ users which are uniformly and randomly distributed within a single cell of radius $500$ meters. The UAV location coordinates $(x_0,y_0)$, the estimated user location coordinates $(\overline{x}_k,\overline{y}_k) $, and the estimated AoD between the UAV and user $k$, $\overline{\theta}_k$, are known at the UAV. The location uncertainty area of user $k$ is assumed to be a circle with a radius $D_k=20$ meters, unless specified otherwise. For ease of presentation, in the sequel, we define the maximum normalized estimation error of the AoD between the UAV and user $k$ as $\rho_k=\frac{\alpha}{ \left |\mathrm{\theta}_k\right |}$, where $\rho_m=\rho_n$, $\forall m,n\in\mathcal{ K}$. Moreover, we employ the nonlinear AAR model in (\ref{steeringvec2}) for all simulations. We choose $\gamma=0.3$ dB for all results shown as this ensured that the desired SINR $\Gamma_{\mathrm{req}_k}$ is achieved for the proposed scheme in all considered cases. Besides, the results presented in this section are obtained by averaging over 1000 channel realizations.
\par
To illustrate the power savings achieved by the proposed scheme, we compare with two baseline schemes. For baseline scheme 1, we adopt zero-forcing beamforming (ZF-BF) at the UAV such that multiuser interference is avoided at the users. Specifically, based on the estimated AoD, the direction of beamforming vector $\mathbf{w_\mathit{k}}$ for desired user $k$ is fixed and lays in the null space of all the other users' channels. Then, we jointly optimize the UAV $x-y$ coordinates ${\mathbf{r}}'_0$ and the power allocated to $\mathbf{w_\mathit{k}}$ under the SDP formulation subject to constraints ${\overline{\mbox{C2}}\mbox{a}}$, ${\overline{\mbox{C2}}\mbox{b}}$, and $\mbox{C3}$ as in (\ref{OP3}). For baseline scheme 2, we employ maximum ratio transmission (MRT), i.e., we set the beamforming vector as $\mathbf{w_\mathit{k}}=\sqrt{p_k}\mathbf{h_\mathit{k}}\left \| \mathbf{h_\mathit{k}} \right \|_2^{-1}$, where $p_k$ and $\mathbf{h_\mathit{k}}$ are the allocated power and the channel vector of the $k$-th user, respectively. Then, the allocated power $p_k$ and the UAV 2-D positioning vector ${\mathbf{r}}'_0$ are jointly optimized for problem (\ref{OP3}) subject to constraints ${\overline{\mbox{C2}}\mbox{a}}$, ${\overline{\mbox{C2}}\mbox{b}}$, and $\mbox{C3}$. In addition, since for most channel realizations the baseline schemes can not simultaneously fulfill the per-antenna power constraint and the QoS requirements, we omit constraint C1 for both baseline schemes to obtain feasible solutions.
\begin{figure}[t]
 \centering
\includegraphics[width=2.9in]{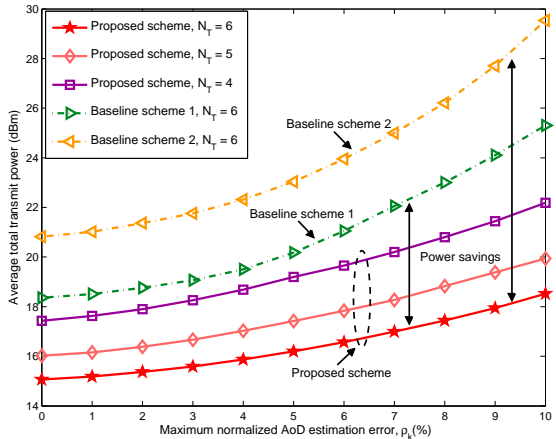} \vspace{-4mm}
\caption{Average total transmit power (dBm) versus the maximum normalized AoD estimation error, $\rho_k$, for different resource allocation schemes with $K=3$ users and minimum required SINRs of $\Gamma_{\mathrm{req}_k}=10$ dB at all users.} \label{fig:power_vs_EE}\vspace{-6mm}
\end{figure}
\par
In Figure \ref{fig:power_vs_EE}, we study the average total transmit power versus the maximum normalized AoD estimation error, $\rho_k$, for $K=3$ users, minimum required user SINRs $\Gamma_{\mathrm{req}_k}=10$ dB, and different numbers of transmit antennas at the UAV, $N_{\mathrm{T}}$. As can be observed, the average total transmit powers for the proposed scheme and the baseline schemes increase monotonically with increasing $\rho_k$. This can be explained by the fact that, as the AoD estimation error increases, it is more difficult for the UAV to perform accurate DL beamforming. Hence, the UAV has to transmit the information signal with higher power to meet the QoS requirements of the users. Moreover, a significant amount of transmit power can be saved by increasing the number of UAV antennas. This is due to the fact that the extra degrees of freedom provided by the additional antennas facilitate a more power efficient resource allocation. On the other hand, the two baseline schemes require a significantly higher total transmit power compared to the proposed scheme. In particular, for the two baseline schemes, the UAV transmitter cannot fully exploit the available degrees of freedom since the beamforming vector $\mathbf{w_\mathit{k}}$ is partially fixed.
\par
\begin{figure}[t]
 \centering 
\includegraphics[width=2.9in]{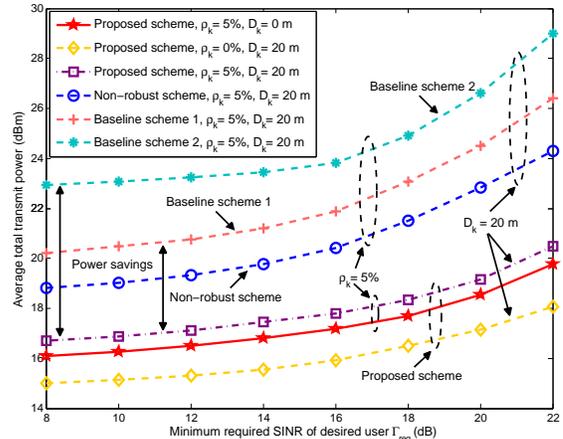} \vspace{-4mm}
\caption{Average total transmit power (dBm) versus minimum required SINR of the users, $\Gamma_{\mathrm{req}_k}$, for different resource allocation schemes with $K=3$ users and $N_{\mathrm{T}}=6$ transmit antennas at the UAV.}\label{fig:power_vs_SINR}\vspace{-5.5mm}
\end{figure}

Figure \ref{fig:power_vs_SINR} illustrates the average total transmit power versus the minimum required user SINRs, $\Gamma_{\mathrm{req}_k}$, for $K=3$ users and different maximum normalized channel estimation errors, $\rho_k$. The UAV has $N_{\mathrm{T}}=6$ transmit antennas. As expected, the average total transmit power of the proposed resource allocation scheme is monotonically nondecreasing with respect to the minimum SINR threshold $\Gamma_{\mathrm{req}}$. This is due to the fact that to meet a larger minimum required SINR in constraint $\mathrm{C2}$, the UAV has to transmit with higher power. Moreover, it can be observed that the total transmit power for the proposed scheme increases with increasing $D_k$. In fact, with increasing user location uncertainty, the UAV is forced to use a less focused beamformer to cover a larger area such that a higher transmit power is needed to satisfy the users' QoS requirements. In Figure \ref{fig:power_vs_SINR}, we also show the average total transmit power of a non-robust scheme. In particular, for the non-robust scheme a similar optimization problem as in (\ref{OP3}) is formulated but the estimated AoD and user location are treated as perfect. Then, using the actual AoDs and user locations, the transmit power allocated to the beamforming vectors $\mathbf{w_\mathit{k}}$ is increased until the QoS requirements of the users are satisfied. Both the non-robust scheme and the two baseline schemes result in a higher total transmit power compared to the proposed robust scheme for the entire considered range of $\Gamma_{\mathrm{req}}$.
\vspace{0mm}
\section{Conclusion}
\vspace{-2mm}
The robust resource allocation design for multiuser DL UAV communication systems was studied in this paper. We formulated the algorithm design as an optimization problem for minimization of the total UAV transmit power taking into account the QoS requirements of the users, the AoD imperfectness caused by UAV jittering, and a user location uncertainty. Thereby, the AAR was linearized with respect to the AoD estimation error. Due to the intractability of the resulting non-convex problem, we transformed it into an equivalent problem by replacing its semi-infinite constraints with LMI constraints. Subsequently, the reformulated problem was solved optimally by employing SDP relaxation. The approximation error introduced by the linearization of the AAR was accounted for by a small increase of the minimum required SINR. Our simulation results revealed dramatic power savings enabled by the proposed robust scheme compared to two baseline schemes. Besides, the robustness of the proposed scheme with respect to UAV jittering and user location uncertainty was confirmed.
\vspace{-1mm}
\section*{Appendix- Proof of Theorem 1}
\vspace{-1mm}
We can verify that the relaxed problem in (\ref{OP3}) is jointly convex with
respect to the optimization variables and the Slater’s
condition is satisfied. As a result, strong duality holds, and the optimal solution of the primal
problem can be obtained by solving the dual problem. Thus, we first write the Lagrangian function of the problem in (\ref{OP3}) in terms of beamforming matrix $\mathbf{W}_\mathit{k}$ as follows:
\vspace{-4.5mm}
\begin{eqnarray}
\hspace{3mm}\mathcal{L}&&\hspace{-6mm}=\underset{ k\in\mathcal{K}}{\sum} \mathrm{Tr}(\mathbf{W}_\mathit{k})+ \sum_{i=1}^{N_{\mathrm{T}}}\xi_i\left[  \underset{k\in\mathcal{K}}{\sum}\mathbf{W}_k\right ]_{i,i}-\underset{ k\in\mathcal{K}}{\sum}\mathrm{Tr}(\mathbf{W}_\mathit{k}\mathbf{Y}_\mathit{k})\notag\\
&&\hspace{-6mm}-\underset{ k\in\mathcal{K}}{\sum}\mathrm{Tr}(\mathbf{S}_{\mathrm{\overline{C2}\mathrm{a}}_{k}}(\mathbf{W}_\mathit{k},\eta_\mathit{k},\mathit{\delta }_\mathit{k})\mathbf{T}_{\mathrm{\overline{C2}\mathrm{a}}_{k}})+\Phi, \label{Lagrangian}
\end{eqnarray}
where $\Phi$ denotes the collection of primal and dual variables and constants that do not affect the proof. The $\xi_i$ are the Lagrange multipliers associated with constraint $\mbox{C1}$. Matrix $\mathbf{T}_{\overline{\mathrm{C}2}\mathrm{a}_{k}}\in \mathbb{C}^{2\times 2}$ is the Lagrange multiplier associated with constraint $\mathrm{\overline{\mbox{C2}}\mathrm{a}}$. Matrix $ \mathbf{Y}_k\in \mathbb{C}^{N_\mathrm{T}\times N_\mathrm{T}}$ is the Lagrange multiplier associated with the positive semidefinite constraint $\mbox{C3}$. Therefore, the dual problem of (\ref{OP3}) is given by
\vspace{-2.5mm}
\begin{equation}
\underset{\mathbf{Y}_k,\mathbf{T}_{\mathrm{\overline{C2}\mathrm{a}}_{k}},\xi_i~}{\mathrm{maximize}~~} \underset{\mathbf{W}_{\mathit{k}}\in\mathbb{H}^{N_\mathrm{T}}}{\mathrm{minimize}}~~ \mathcal{L}(\mathbf{W}_k,\mathbf{Y}_k,\mathbf{T}_{\mathrm{\overline{C2}\mathrm{a}}_{k}},\zeta).\label{DP}   
\end{equation}
Then, we study the structure of the optimal $\mathbf{W}_k$ of dual problem (\ref{OP3}) based on the Karush-Kuhn-Tucker (KKT) conditions. The KKT conditions for the optimal $\mathbf{W}_k^*$ are given by
\vspace{-2.5mm}
\begin{eqnarray}
\xi_i^*\geq 0, \mathbf{Y}_k^*,\mathbf{T}^*_{\mathrm{\overline{C2}\mathrm{a}}_{k}}&&\hspace{-6mm}\succeq \mathbf{0},\label{KKTSP1}\\    
\mathbf{Y}_k^*\mathbf{W}_k^*&&\hspace{-6mm}=\mathbf{0},\label{KKTSP2}\\
\triangledown_{\mathbf{W}_k^*}\mathcal{L}&&\hspace{-6mm}=\mathbf{0}, \label{grasp}
\end{eqnarray}
where $\xi_i^*$, $\mathbf{T}^*_{\mathrm{\overline{C2}\mathrm{a}}_{k}}$, and $\mathbf{Y}_k^*$ are the optimal dual variables for dual problem ($\ref{DP}$), and $\triangledown_{\mathbf{W}_k^*}\mathcal{L}$ denotes the gradient of the Lagrangian function with respect to $\mathbf{W}_k^*$. Moreover, we obtain from the KKT condition in ($\ref{grasp}$)
\vspace{-3mm}
\begin{equation}
\mathbf{Y}_k^*=\mathbf{I}_{N_{\mathrm{T}}}-\mathbf{\Delta},    \label{KKTSPex}\vspace{-2mm}
\end{equation}
where
\vspace{-3mm}
\begin{align}
\hspace{-3mm}\mathbf{\Delta} =\mathbf{U}_{\mathbf{a}_k}\mathbf{T}^*_{\mathrm{\overline{C2}\mathrm{a}}_{k}}\mathbf{U}_{\mathbf{a}_k}^H-\hspace{-2mm}\underset{ r\in\mathcal{K}\setminus \left \{ k \right \}}{\sum}\hspace{-2mm}\Gamma _{\mathrm{req_{\mathit{r}}}} \mathbf{U}_{\mathbf{a}_r}\mathbf{T}^*_{\mathrm{\overline{C2}\mathrm{a}}_{r}}\mathbf{U}_{\mathbf{a}_r}^H-\mathbf{\Xi}^*, 
\end{align}
and $\mathbf{\Xi}^*$ is defined as $\mathbf{\Xi}^*\overset{\Delta}{=}\mathrm{diag}( \xi_1^*,\cdots , \xi_{N_{\mathrm{T}}}^*)$.
\par
Next, we reveal that $\mathbf{\Delta}$ is a positive semidefinite matrix by contradiction. Specifically, if $\mathbf{\Delta}$ is a negative definite matrix, then from ($\ref{KKTSPex}$), $\mathbf{Y}_k^*$ must be a full-rank positive definite matrix. Considering the KKT condition in ($\ref{KKTSP2}$), this implies $\mathbf{W}_k^*=\mathbf{0}$ which cannot to be the optimal solution for $\Gamma _{\mathrm{req}_k}>0$. Therefore, we focus on the case where $\mathbf{\Delta}$ is a positive semidefinite matrix in the rest of the proof. Due to the KKT condition in ($\ref{KKTSP1}$), which indicates that matrix $\mathbf{Y}_k^*=\mathbf{I}_{N_{\mathrm{T}}}-\mathbf{\Delta}$ is also positive semidefinite, we have
\vspace{-3mm}
\begin{equation}
1\geq \nu ^{\mathrm{max}}_{\mathbf{\Delta}}\geq 0,    
\end{equation}
where $\nu ^{\mathrm{max}}_{\mathbf{\Delta}}\in\mathbb{R}$ denotes the maximum eigenvalue of matrix $\mathbf{\Delta}$. Reviewing the KKT condition in ($\ref{KKTSPex}$), for the case where $1> \nu ^{\mathrm{max}}_{\mathbf{\Delta}}$, we can see that matrix $\mathbf{Y}_k^*$ turns into a positive definite matrix with full rank. Again, this leads to $\mathbf{W}_k^*=\mathbf{0}$ which contradicts the positive minimum required SINR $\Gamma _{\mathrm{req}_k}>0$. Thus, for the optimal solution, the maximum eigenvalue of matrix $\mathbf{\Delta}$ must fulfill $\nu ^{\mathrm{max}}_{\mathbf{\Delta}}=1$. Since the users are randomly distributed within UAV's service area, the case where multiple eigenvalues have the same value $\nu ^{\mathrm{max}}_{\mathbf{\Delta}}$ occurs with probability zero. Hence, we focus on the case where $\mathbf{\Delta}$ has a unique maximum eigenvalue which leads to $\mathrm{Rank}(\mathbf{Y}_k^*)=N_\mathrm{T}-1$. Moreover, in order to obtain a bounded optimal dual solution, we span the null space of $\mathbf{Y}_k^*$ by a vector $\mathbf{e}_{\mathbf{\Delta}}^{\mathrm{max}}$, i.e., $\mathbf{Y}_k^*\mathbf{e}_{\mathbf{\Delta}}^{\mathrm{max}}=\mathbf{0}$, where $\mathbf{e}_{\mathbf{\Delta}}^{\mathrm{max}}\in \mathbb{C}^{N_{\mathrm{T}}\times 1}$ is the unit-norm eigenvector of matrix $\mathbf{\Delta}$ corresponding to the maximum eigenvalue $\nu ^{\mathrm{max}}_{\mathbf{\Delta}}$. As a result, for $\Gamma _{\mathrm{req}_k}>0$, the optimal beamforming matrix $\mathbf{W}_k^*$ satisfies $\mathrm{Rank}(\mathbf{W}_k^*)=1$ and can be expressed as
\vspace{-2.5mm}
\begin{equation}
\mathbf{W}_k^*= \beta \mathbf{e}_{\mathbf{\Delta}}^{\mathrm{max}}(\mathbf{e}_{\mathbf{\Delta}}^{\mathrm{max}})^H,
\end{equation}
where $\beta$ is a parameter which guarantees that the per-antenna transmit power satisfies constraint $\mbox{C1}$. \qed

\vspace{-2mm}
\bibliographystyle{IEEEtran}

\end{document}